# Molecular chaperones:
# The modular evolution of cellular networks

**Tamás Korcsmáros, István A. Kovács, Máté S. Szalay and Péter Csermely***
Department of Medical Chemistry, Semmelweis University, Budapest, Hungary

*Corresponding Author: Peter Csermely – Department of Medical Chemistry, Semmelweis University, Puskin street 9., H-1088 Budapest, Hungary, phone: +36-1-266-2755, extn: 4102, FAX: +36-1-266-6550, Email: csermely@puskin.sote.hu

**Abstract** Molecular chaperones play a prominent role in signaling and transcriptional regulatory networks of the cell. Recent advances uncovered that chaperones act as genetic buffers stabilizing the phenotype of various cells and organisms and may serve as potential regulators of evolvability. Chaperones have weak links, connect hubs, are in the overlaps of network modules and may uncouple these modules during stress, which gives an additional protection for the cell at the network-level. Moreover, after stress chaperones are essential to re-build inter-modular contacts by their low affinity sampling of the potential interaction partners in different modules. This opens the way to the chaperone-regulated modular evolution of cellular networks, and helps us to design novel therapeutic and anti-aging strategies.

## 1. Introduction: cellular networks and chaperones

Most of the molecular interactions of our cells, like the self-association of lipids to membranes, are rather unspecific, and can be described in general terms. However, a relatively restricted number of interactions between cellular molecules have a high affinity, are unique and specific, and require a network approach for a better understanding and prediction of their changes after various environmental changes, like stress (Albert, 2005; Barabasi and Oltvai, 2004; Boccaletti et al., 2006; Csermely, 2006). The protein-protein interaction network is a good example for the network description of unique cellular interactions between molecules (Fig. 1). Here the elements of the network are proteins, and the links between them are permanent or transient bonds (Gavin et al., 2006; von Mering et al, 2002; Rual et al, 2005). Cytoskeletal filaments form the elements of the cytoskeletal network, and the bonds between them are the links. Membrane segments (membrane vesicles, domains, rafts, of cellular membranes) and cellular organelles (mitochondria, lysosomes, segments of the endoplasmic reticulum, etc.) are the elements of the membranous, organellar network, and they are linked by protein complexes and/or membrane channels. Both membranes and organelles contain large protein-protein interaction networks (Aon et al., 2006). In signaling networks the elements are proteins or protein complexes and the links are highly specific interactions between them, which undergo a profound change (either activation or inhibition), when a specific signal reaches the cell (White and Anderson, 2005). In metabolic networks the network elements are metabolites, such as glucose, or adenine, and the links between them are the enzyme reactions, which transform one metabolite from the other (Pal et al, 2006). Finally, gene transcription networks have two types of elements, transcriptional factor-complexes and the DNA gene sequences, which they regulate. Here the transcriptional factor-complexes may initiate or block the transcription of the messenger RNA-s. The links between these elements are the functional (and physical) interactions between the proteins (sometimes RNA-s) and various parts of the gene sequences in the cellular DNA (Yu and Gerstein, 2006).

Cellular networks often form small worlds, where two elements of the network are separated by only a few other elements. Networks of our cells usually have a scale-free degree distribution, which means that these networks have hubs, i.e. elements, which have a large number of neighbors. These networks are rich in motifs, which are regularly appearing combinations of a few adjacent network elements, and contain hierarchical modules, or in other words: are forming hierarchical communities (Albert, 2005; Barabasi and Oltvai, 2004; Boccaletti et al., 2006; Csermely, 2006). The complex architecture of cellular networks solves four major tasks (Fig. 2). (1) The first task is the local dissipation of the perturbations/noise coming from outside the cell, and from the stochasticity of intracellular events. (2) The second task is the efficient and reliable global transmission of signals from one distant element of the cell to another. (3) The third task is the discrimination between signals and noise via the continuous remodeling of these networks during the evolutionary learning process of the cell. (4) The fourth task is the protection against the continuous random damage of free radicals and other harmful effects during stress and aging. During the execution of these tasks the assembly of network elements produces a vast number of emergent properties of networks, which can only be understood, if we study the whole network and cannot be predicted knowing the behavior of any of its elements (Csermely, 2006).

The complex topology of cellular networks was evolved to solve these four tasks. As a relatively simplified view, hubs help to confine most of the perturbations to a local environment, while the small world character allows the global propagation of signals. Motifs and hierarchical modules help both the discrimination between the two, and provide stability at the network level (which is helped by a number of repair functions at the molecular level; Csermely, 2006). However, this summary of the major features of cellular networks is largely a generalization, and needs to be validated through critical scrutiny of the datasets, sampling procedures and methods of data analysis at each network examined (Arita, 2004; Tanaka et al., 2005).

**2. Chaperones in cellular networks**
Molecular chaperones bind and release a large number of damaged proteins, which requires a large promiscuity in their interactions. Not surprisingly, chaperones form low affinity, dynamic temporary interactions (weak links) in cellular networks (Table 1; Csermely, 2004; Tsigelny and Nigam, 2004). Chaperones have a large number of neighbors acting like hubs in the yeast protein-protein interaction network (Table 2). The large proportion of hubs in the close vicinity of chaperones gives a central position to these proteins in the protein-protein network, which may help the chaperone-mediated cross talk between pathways. Most importantly, chaperones are inter-modular elements of cellular networks, assembling the modular structure of the cell (Table 3).

De-coupling of network elements and modules is a generally used method to stop the propagation of damage (Albert, 2005; Barabasi and Oltvai, 2004; Boccaletti et al., 2006; Csermely, 2006). When the cell experiences stress, chaperones become increasingly occupied by damaged proteins. The chaperone overload, a usual phenomenon in sick or aging cells (Csermely, 2001; Nardai et al, 2002), meaning the competitive inhibition of chaperones, can be experienced due to the large number of equally low affinity interactions of these proteins. Chaperone inhibition might lead to an 'automatic' de-coupling of all chaperone-mediated networks providing an additional safety measure for the cell (Soti et al., 2005a). However, the really important task comes after the stress is over: the cell has to re-assemble the de-coupled modules again. This again is helped by the newly synthesized molecular chaperones, since their low affinity interactions effectively sample a large number of proteins in different modules and allow to bind these modules each other in a very flexible, partially stochastic manner. This gives the cell a refined and flexible way for the gradual build-up of the complex modular structure and function, when the stress is already over (Fig. 3).

**3. Chaperone-regulated modular evolution of cellular networks**
One of the best examples of chaperone-mediated emergent network properties was shown by Susannah Rutherford and Susan Lindquist, when they discovered that Hsp90 acts as a buffer to conceal the phenotype of the genetic changes in *Drosophila melanogaster* (Rutherford and Lindquist, 1998). Chaperone-induced genetic buffering is released upon stress, which causes the sudden appearance of the phenotype of previously hidden mutations, helps population survival and gives a possible molecular mechanism for fast evolutionary changes. On the other hand, stress-induced appearance of genetic variation at the level of the phenotype cleanses the genome of the population by allowing the exposure and gradual disappearance of disadvantageous mutations by natural selection. After the initial report of Rutherford and Lindquist (1998), the effect was extended to other chaperones and to *Escherichia coli*,

*Arabidopsis thaliana* and the evolution of resistance in fungi (Cowen and Lindquist, 2005; Fares et al., 2002; Queitsch et al., 2002). In recent years the scientific community has became increasingly aware of the idea that not only chaperones but a large number of other proteins may also regulate the diversity of the phenotype (Bergman and Siegal, 2003; Csermely, 2004; 2006). If a general explanation is sought, it is more likely to be related to the network properties of the cell. In this context, the weak links of chaperones, their central position linking hubs to each other and their inter-modular links may all help their regulatory role in the evolvability of complex systems.

The remodeling of the inter-modular contacts is an especially intriguing idea for the explanation of chaperone-mediated sudden changes in the emergent properties (such as the phenotype) of cellular networks. Different assembly of slightly changed cellular modules may cause profound and abrupt changes of the functional repertoire without major changes of the underlying structure of protein-protein interactions. This gives an exploratory but stable mechanism for the evolution of cellular networks.

As a summary we may say that as the system grows more complex, due to its weak links and modular topology it has higher chances to stabilize itself and its environment. Complexity and increased stability would lead to a restricted chance for evolution. However, very importantly, as a consequence of the very same weak links and modules, the complex system acquires a higher evolvability, which helps to preserve its chances for development in spite of the stricter requirements to accomplish this task.

## 4. Chaperone therapies

Cellular networks are remodeled in various diseases and after stress. Proper interventions to push the equilibrium towards the original state may not be limited to single-target drugs, which have a well-designed, high affinity interaction with one of the cellular proteins. In agreement with this general assumption, several examples show that multi-target therapy may be superior to the usual single-target approach. The best known examples of multi-target drugs include Aspirin, Metformin or Gleevec as well as combinatorial therapy and natural remedies, such as herbal teas (Csermely et al., 2005). Due to the multiple regulatory roles of chaperones, chaperone-modulators provide additional examples for multi-target drugs. Indeed, chaperone substitution (in the form of chemical chaperones), the help of chaperone induction and chaperone inhibition are all promising therapeutic strategies (Bernier et al, 2004; Neckers and Neckers, 2005; Soti et al., 2005b; Vigh et al., 1997).

## 5. Discussion

Chaperones regulate cellular functions at two levels. In several cases they interact with a specific target protein, and become mandatory to its folding as well as for the assistance in the formation of specific protein complexes (and in the prevention of the assembly of others). These specific interactions make chaperones important parts of the core of cellular networks, such as the protein net, the signaling network, the membranous and organellar network as well as the transcriptional network. However, in most cases chaperones have only a low affinity, temporary interactions, i.e. 'weak links' with most of their targets. Importantly, many of these weakly bound partners are in different modules of the cell. Thus, the inhibition of chaperone-mediated weak links might not only lead to a rise in cellular noise, but efficiently de-couples the whole network. By this complex version of the 'error-catastrophe', chaperone inhibitors help us to combat cancer. In contrast, chaperone activation may decrease cellular noise, stabilize and integrate cells and thus give a general aid against aging and diseases. The chaperone-mediated disassembly and re-assembly of the modules of cellular networks gives a very efficient tool for engineering profound changes in the emergent properties (functions) of cellular networks with subtle changes at the molecular level. Moreover, these changes are partially reversible, which makes a safety-measure, if the invention was not a successful step giving an increased fitness. Chaperones emerge as key regulators of the evolvability of cellular networks in a modular fashion. The assessment of the multiple roles of chaperones in the context of cellular networks is only just beginning.

*Acknowledgements*
Work in the authors' laboratory was supported by research grants from the EU (FP6506850, FP6-016003) and from the Hungarian National Research Initiative (1A/056/2004 and KKK-0015/3.0).

**Table 1. Chaperones connect hubs in the yeast protein-protein interaction network**

| Chaperone class | % of hub Neighbors |
|---|---|
| Hsp70 | 41 |
| Hsp90 | 49 |
| average | 28 |

The ratio of neighboring hubs (proteins having more than a 100 interacting partners) to all neighbors for the Hsp70 and Hsp90 chaperone classes (containing all isoforms and their co-chaperones) was calculated using the annotated yeast protein-protein interaction database of von Mering et al (2002). The skewed distribution of data and the large amount of low-probability interactions, which include potential artifacts, makes the statistical evaluation of the differences between the chaperone class values and the average values difficult.

**Table 2. Percent of high confidence chaperone interactions in two yeast protein-protein interaction networks**

| Chaperone class | Annotated yeast database | Affinity purified yeast database |
|---|---|---|
| Hsp60 | 1.2 | 3.7 |
| Hsp70 | 0 | 0.7 |
| Hsp90 | 0 | 1.4 |
| average | 4.0 | 7.5 |

The percentage of high confidence partners from all links of the respective chaperone classes (containing all isoforms and their co-chaperones) were calculated using the annotated and affinity purified yeast protein-protein interaction databases of von Mering et al (2002) and Gavin et al. (2005), respectively. Low confidence interactions may contain a considerable amount of artifacts, but are also enriched in low affinity chaperone-neighbor links. The average is the average percentage of high confidence neighbors of all proteins in the database. The skewed distribution of data and the large amount of low-probability interactions, which include potential artifacts, makes the statistical evaluation of the differences between the chaperone class values and the average values difficult.

**Table 3. Number of chaperone-connected modules in protein-protein interaction networks**

| Chaperone class | Annotated yeast database | Affinity purified yeast database | Human affinity- and literature-based database |
|---|---|---|---|
| Mt-Hsp60 | 0 | 49 | 5 |
| Hsp70 | **2** (SSA4) | **118** (SSA1, SSA3, SSA4) | **10** (HSPA1A) |
| Hsc70 | 0 | **116** (SSB1, SSB2) | **10** (HSPA8) |
| Hsp110 | **2** (SSE2) | **46** (SSE1) | **5** (HSPH1) |
| Hsp90 | **2** (Hsp82) | **63** (Hsc82+ Hsp82) | **20** (HSPCA) |
| References and number of modules | von Mering et al, 2002 (26 modules total; Valente and Cusick, 2006) | Gavin et al., 2006 (147 modules total) | Rual et al. 2005 (173 modules of total) |

The numbers of chaperone-connected modules of the respective chaperone classes (containing all isoforms) were calculated using the annotated yeast, affinity purified yeast and human protein-protein interaction databases of von Mering et al (2002), Gavin et al. (2006) and Rual et al. (2005), respectively. Protein modules were obtained from the original database publications or from Valente and Cusick (2006) in case of the annotated yeast database. The low number of connected modules in the annotated yeast database reflects the fact, that the module determination method of Valente and Cusick (2006) used only the high affinity interactions resulting in a small number of modules only, while the low number of connected modules in the human database is due to the poor representation of chaperones in this database. The symbols in parentheses refer to the most important inter-modular chaperones.

**Figures**

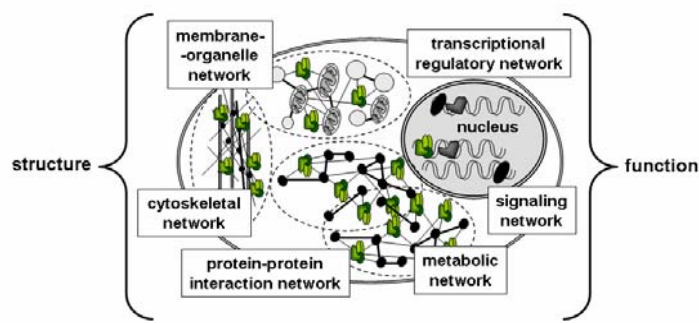

**Figure 1. Cellular networks.** The figure illustrates the most important networks in our cells. The protein-protein interaction network, the cytoskeletal network and the membranous, organellar networks provide a general scaffold of the cell containing the physical interactions between cellular proteins. Other networks, like the signaling, transcriptional, or metabolic networks are functionally defined. In the signaling network elements of various signaling pathways are linked by the interactions between them. In the transcriptional regulatory network the elements are the transcription factors, genes and the connecting links are functional interactions between them. In the metabolic network we have the various metabolites as elements and the enzyme reactions as links. All these networks highly overlap with each other, and some of them contain modules of other networks.

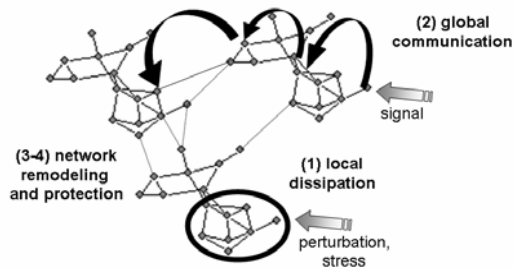

**Figure 2. Major tasks of cellular networks.** (1) Local dissipation of the perturbations/noise coming from outside the cell, and from the stochastic elements of intracellular reactions. (2) Efficient and reliable global transmission of signals from one element of the cell to another. (3) Discrimination between signals and noise via the continuous remodeling of these networks during the evolutionary learning process of the cell. (4) Protection against the continuous random damage of free radicals and other harmful effects during stress and aging.

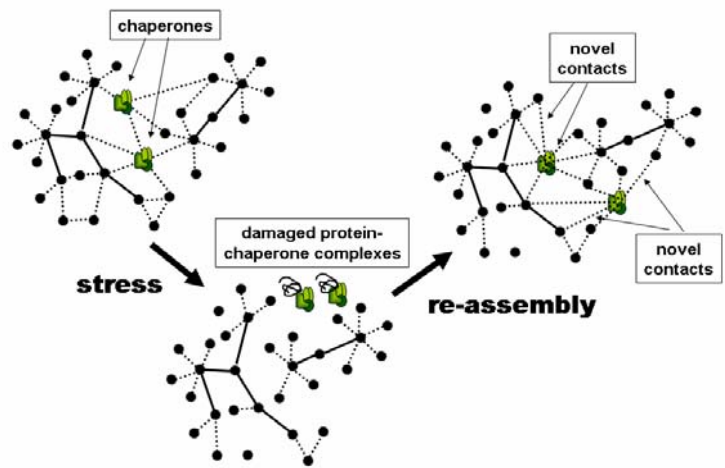

**Figure 3. Chaperone-mediated modular evolution of networks.** Chaperones link a large number of modules in the cellular networks. Occupancy and inhibition of chaperones (e.g. during stress) efficiently de-couples these modules. After the stress is over, modules start to re-assemble, which is helped by the partially random contacts of original and *de novo* synthesized molecular chaperones between modules. Subtle changes in modular reassembly may lead to profound changes of the emergent properties (functions) of the whole network. Importantly, the mis-assembly of modules is reversible. This expands the flexibility of the system, and helps its adaptation to the changing environment.